# Single-shot quantitative differential phase contrast imaging combined with programmable polarization multiplexing illumination

Siying Liu,[1] Chuanjian Zheng,[1] Qun Hao,[1,2,4] Xin Li,[3] and Shaohui Zhang[1*]

[1]School of Optics and Photonics, Beijing Institute of Technology, Beijing 100081, China
[2]Changchun University of Science and Technology, Changchun 130022, China
[3]Department of General Surgery, Xiangya Hospital, Central South University, Changsha 410011, China
[4]e-mail: qhao@bit.edu.cn
*Corresponding author: zhangshaohui@bit.edu.cn

**We propose a single-shot quantitative differential phase contrast (DPC) method with polarization multiplexing illumination. In the illumination module of our system, the programmable LED array is divided into four quadrants and covered with polarizing films of four different polarization angles. We use a polarization camera with polarizers before the pixels in the imaging module. By matching the polarization angle between the polarizing films over the custom LED array and the polarizers in the camera, two sets of asymmetric illumination acquisition images can be calculated from a single-shot acquisition image. Combined with the phase transfer function, we can calculate the quantitative phase of the sample. We present the design, implementation, and experimental image data demonstrating the ability of our method to obtain quantitative phase images of the phase resolution target, as well as Hela cells.**

As a label- and stain-free method to observe the internal structure of weakly scattering samples, phase imaging has a wide range of applications in biological, pathological, and material research [1]. Compared to qualitative phase imaging methods such as phase contrast and differential interference contrast (DIC), quantitative phase imaging (QPI) can quantify the optical path length of the sample by deconvolution or iteration algorithms [2], making it an important tool for quantitative studies in multidisciplinary fields, such as the observation of nanoscale electrophysiological phenomena [3] and the interaction between normal cells and diseased cells [4]. QPI can be classified as interferometric and noninterferometric methods depending on the imaging principle. The interferometric methods are easily affected by environmental noise due to the interference principle. On the contrary, the noninterferometric QPI methods such as transport of intensity equation (TIE) [5,6], Fourier ptychographic microscopy (FPM) [7,8], and differential phase contrast (DPC) [9,10] can alleviate this problem and provide higher robustness [11]. Among the above methods, DPC is a promising tool for observing live cells with high imaging efficiency and relatively easy algorithm.

In DPC, phase contrast images are calculated by four intensity images taken with two groups of axial orthogonal asymmetric complementary illumination, which are linked to the quantitative phase of the sample through the phase transfer function [12]. However, the acquisition strategy for multiple measurements reduces the temporal resolution of DPC, and the measurement of highly dynamic images under this strategy is subject to artifact problems. Therefore, the programmable LED array changing illumination modes rapidly by coding is introduced into the DPC system to realize the real-time DPC [12,13]. Subsequently, single-shot quantitative DPC emerged, relying on color light sources and color cameras for wavelength multiplexing. Although color light sources have evolved from the programmable LED array [14] to custom filters [15], they can't avoid the problem of color leakage, which need to introduce color verification and correction methods after raw data acquisition.

We proposed a single-shot quantitative DPC imaging method with programmable polarization multiplexing illumination. The proposed method is implemented on a programmable LED array microscope platform, which can generate various illumination modes using computer software via USB with an STM32 microcontroller. Fig.1 depicts our DPC imaging system. The microscopic platform uses the custom programmable LED array instead of the illumination module of a conventional commercial microscope. The LED array (5 mm pitch, 22×22 RGB LED matrix panel) is divided into four quadrants with the center as the origin, covering 0°, 45°, 90° and 135° polarization angles of the polarizing films (extinction ratio: >500:1, 400-450nm; >5000:1 450-700nm) respectively. The polarized light, mixed at four polarization angles emitted from the illumination module, first passes through the sample, the standard microscope's objective (0.5NA, 20×, 0.6NA,40×), tube lenses and is subsequently received by the polarization camera (BFS-U3-51S5, 2448×2048 pixels, 3.45$\mu m$).

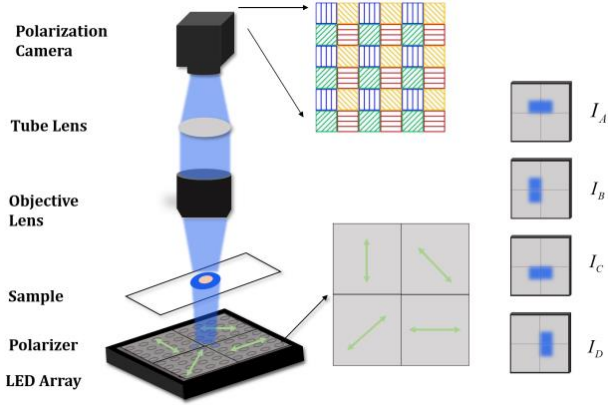

Fig.1. Schematic diagram of the proposed system. The illumination module includes the custom programmable LED array and the polarizing films composed of four polarizers with different polarization angles, and the same four angles of polarizers distributed in front of the pixel sensor of the polarization camera in the imaging module, combining two polarizers in the system we can calculate the images under orthogonal asymmetric illumination modes ($I_A, I_B, I_C, I_D$).

The intensity image of the sample acquired by the camera under a single LED illumination can be written as:

$$I(r) = \left| \Gamma\left\{ \Gamma\left[ S(u,r) \cdot o(r) \right] \cdot P(u) \right\} \right|^2, \quad (1)$$

where $\Gamma$ denotes the Fourier transfer function. $u, r$ denotes spatial coordinates $(x, y)$. $I$ is the intensity of the camera. $P$ is the pupil function in the objective lens, and $o$ denotes the sample's transmittance function. The light from each LED of the illumination module can be written as a plane wave since its distance from the sample is sufficiently far:

$$S(u,r) = \sqrt{g(r)} \exp(i2\pi u \cdot r), \quad (2)$$

where $g$ denotes the intensity of the light source, the intensity image acquired under non-coherent illumination of multiple LEDs can be written as an integral of each LED illumination result:

$$I(r) = \iint \left| \Gamma\left\{ \Gamma\left[ S(u,r) \cdot o(r) \right] \cdot P(u) \right\} \right|^2 d^2u, \quad (3)$$

Under a weak object approximation, the intensity of a weakly scattering sample in the Fourier space can be written as:

$$\tilde{I}(u) = B\delta(u) + H_{abs}(u) \cdot \tilde{\mu}(u) + H_{ph}(u) \cdot \tilde{\phi}(u), \quad (4)$$

where $u$ denotes the spatial frequency coordinates. $\tilde{\mu}$ and $\tilde{\phi}$ denote the sample's absorption and phase in the spectrum, respectively. $B, H_{abs}$ and $H_{ph}$ are coefficients the of background, absorption and phase term, and the specific expressions are given by [12].

In our system, since the illumination module incorporates the polarizing films, the intensity image of the sample under non-coherent illumination acquired by the camera can be written as a sum of four integral parts:

$$I(r) = \iint_{D_0} \left| \Gamma\left\{ \Gamma\left[ \sqrt{p_0 g(r)} \exp(i2\pi u \cdot r) \cdot o(r) \right] \cdot P(u) \right\} \right|^2 du^2 +$$
$$\iint_{D_{45}} \left| \Gamma\left\{ \Gamma\left[ \sqrt{p_{45} g(r)} \exp(i2\pi u \cdot r) \cdot o(r) \right] \cdot P(u) \right\} \right|^2 du^2 + \quad (5)$$
$$\iint_{D_{90}} \left| \Gamma\left\{ \Gamma\left[ \sqrt{p_{90} g(r)} \exp(i2\pi u \cdot r) \cdot o(r) \right] \cdot P(u) \right\} \right|^2 du^2 +$$
$$\iint_{D_{135}} \left| \Gamma\left\{ \Gamma\left[ \sqrt{p_{135} g(r)} \exp(i2\pi u \cdot r) \cdot o(r) \right] \cdot P(u) \right\} \right|^2 du^2,$$

where, $p_j (j = 0, 45, 90, 135)$ is the transmittance coefficient of the polarizing film at the corresponding angle. $D_j (j = 0, 45, 90, 135)$ in the integral term denotes the LED area under the corresponding angle polarizing films. Under the isotropic sample assumption, for the non-polarized light emitted by the LED array, the polarizing coefficient $p$ of the illumination module can be regarded as a constant. Thus, Eq. (4) can be written as follows:

$$I(r) = p_0 \cdot I_{p0} + p_{45} \cdot I_{p45} + p_{90} \cdot I_{p90} + p_{135} \cdot I_{p135}, \quad (6)$$

$$I_{pj} = \iint_{D_j} \left| \Gamma\left\{ \Gamma\left[ \sqrt{g(r)} \exp(i2\pi u \cdot r) \cdot o(r) \right] \cdot P(u) \right\} \right|^2 du^2$$
$$(j = 0, 45, 90, 135), \quad (7)$$

where $I_{pj}$ denotes the intensity at the camera when LEDs located in corresponding angle area illuminate the sample without polarizing films, significantly, the polarization camera used in our system is distributed with four kinds of polarizers, so it is necessary to add the polarizing coefficient before the resultant intensity captured on the camera, that is to say, the four polarization images calculated from the intensity image received by the polarization camera can be written as:

$$I_j(r) = p_j \cdot \left( p_0 \cdot I_{p0} + p_{45} \cdot I_{p45} + p_{90} \cdot I_{p90} + p_{135} \cdot I_{p135} \right) \quad (8)$$
$$(j = 0, 45, 90, 135),$$

Marius law gives the relationship between the transmission coefficients of two linear polarizers in the optical system:

$$p_\xi \cdot p_\eta = \cos^2(\xi - \eta), \quad (9)$$

where $\xi, \eta$ are the polarization angles of the polarizers, respectively. Substitute Eq. (9) into Eq. (8), we can obtain:

$$\begin{bmatrix} I_0 \\ I_{45} \\ I_{90} \\ I_{135} \end{bmatrix} = \begin{bmatrix} 1 & 1/2 & 0 & 1/2 \\ 1/2 & 1 & 1/2 & 0 \\ 0 & 1/2 & 1 & 1/2 \\ 1/2 & 0 & 1/2 & 1 \end{bmatrix} \begin{bmatrix} I_{P0} \\ I_{P45} \\ I_{P90} \\ I_{P135} \end{bmatrix}, \quad (10)$$

where $I_j (j=0,45,90,135)$ denotes the four polarization images calculated from the intensity image received by the polarization camera. Through mathematical derivation, the equations of sample intensity images under four illumination modes shown in Figure 1 can be obtained:

$$\begin{bmatrix} I_A \\ I_B \\ I_C \\ I_D \end{bmatrix} = \begin{bmatrix} I_{p90}+I_{P135} \\ I_{p90}+I_{P45} \\ I_{p45}+I_{P0} \\ I_{p135}+I_{P0} \end{bmatrix} = \begin{bmatrix} -1/4 & -1/4 & 3/4 & 3/4 \\ -1/4 & 3/4 & 3/4 & -1/4 \\ 3/4 & 3/4 & -1/4 & -1/4 \\ 3/4 & -1/4 & -1/4 & 3/4 \end{bmatrix} \begin{bmatrix} I_0 \\ I_{45} \\ I_{90} \\ I_{135} \end{bmatrix}, \quad (11)$$

The intensity of DPC is defined as:

$$I^{DPC}(r) = \frac{I_{A/B}(r) - I_{C/D}(r)}{I_{A/B}(r) + I_{C/D}(r)}, \quad (12)$$

Quantitative phase information can then be solved using the Tikhonov regulation:

$$\phi_{tik}(r) = \Gamma^{-1} \left\{ \frac{\sum_j H^*_{ph,j}(u) \cdot \Gamma\left[I_{DPC,j}(u)\right]}{\sum_j |H_{ph,j}(u)|^2 + \alpha} \right\}, \quad (13)$$

where $\alpha$ presents the regulation parameter, the index $j$ denotes the DPC measures along $j-th$ axis.

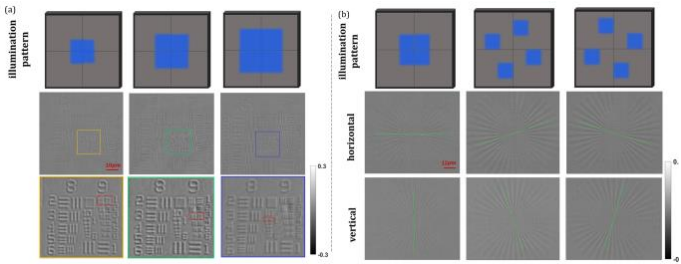

Fig. 2. Qualitative phase results of the phase resolution target under multi-mode illumination. (a) Experimental results of the USAF Target under the resolution lifting illumination. (b) Experimental results of "Focus Star" under the multi-axial phase contrast illumination.

In this paper, the system uses the programmable LED array as illumination, which can provide multi-mode illumination flexibly. For the phase contrast images calculated by Eq. (12), we give two illumination modes of resolution lifting and multi-axial phase contrast and verify them using the phase plate's absorptive USAF target and 'Focus Star'. The phase contrast images of the USAF target under the light source with different illumination NA are given in Fig.2(a). The result is that the phase contrast images obtained with a 4×4 LED array centered on the LED coordinate origin can reach USAF target group 9 Element 1, and the light source of a 6×6 led array in the same position can reach USAF target group 9 Element 3. A 10×10 led array can reach USAF target group 10 Element 1. Experimental results show that the image resolution increases with the increase of the NA in the system. At the same time, we also verified the phase contrast images for other axes centered at the origin of the coordinates. The "focus star" results in the phase plate acquired by the 20x objective are given in Fig. 2(b), and the horizontal symmetry axis of three orthogonal axial phase contrast images are rotated by 0°, 16°, and -16 °, respectively. The results show that the higher contrast directions in the phase contrast images are the same as the selected axes for the asymmetric illumination pattern.

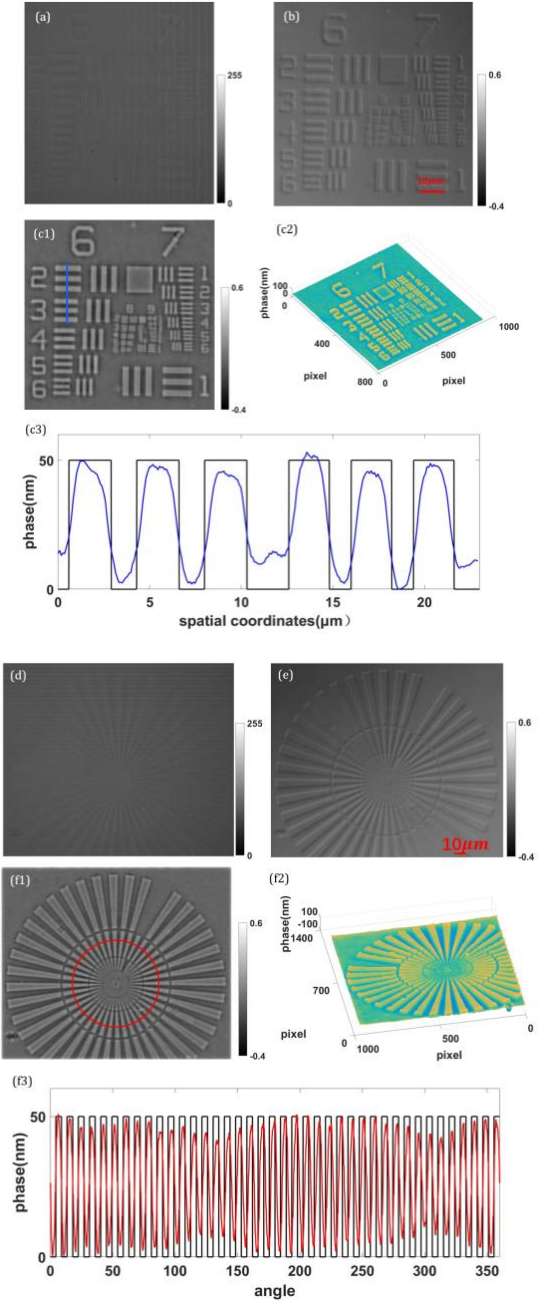

Fig. 3. Quantitative phase results of the phase resolution target under multi-mode illumination. (a) and (d) Captured images of the USAF Target and "Focus Star" using our single-shot method. (b) and (e) Qualitative phase contrast images. (c1) and (c1) Quantitative phase images. (c2) and (f2) 3D pseudo-color morphological distribution. (c3) Phase values along the blue line in (c1). (f3) Phase values along the red circles in (f1).

We verified the quantitative effect of the proposed single-shot DPC using the phase resolution target (QPT). Under the annular illumination (inner radius is $0.75 NA_{objective}$, outer radius is $1.25 NA_{objective}$) at 437 nm wavelength provided by the programmable LED, Figs. 3(a) and 3(d) give the images captured by the polarization camera, which have low phase contrast. The phase contrast images shown in Fig. 3(b) and 3(e) improve image contrast but still do not give the sample quantitative phase information. Figs. 3(c1) and 3(f1) show the quantitative phase calculated by Eq. (13). Figs. 3(c2) and 3(f2) give 3D pseudo-color morphological distribution of the samples according to the measured phase. We extract the phase values of the blue lines shown in Fig. 3(c1) to plot the quantified height curves. As shown in Fig. 3(c3), we can see that the height of the sample calculated from the measured phase information is almost consistent with its true height. Similarly, Fig. 3(f3) gives the height distribution of the red circle area in Figure 3(f1), which shows that our single-shot illumination almost achieves isotropic Phase reconstruction.

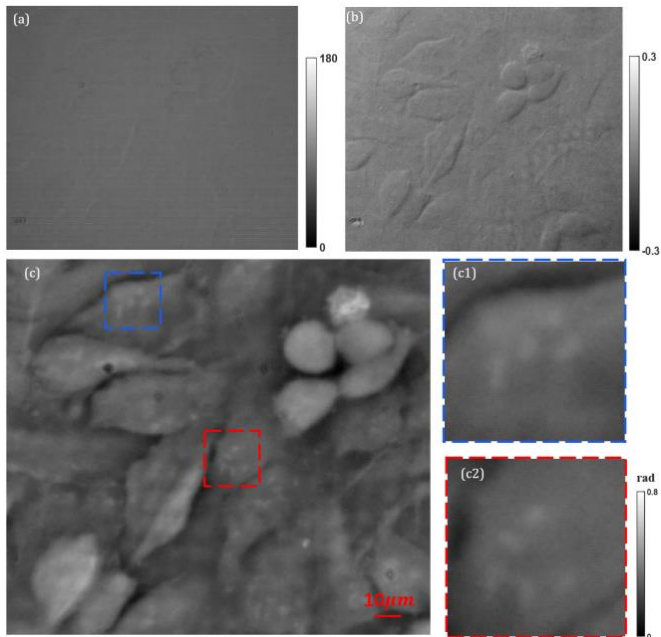

Fig. 4. Quantitative phase images of Hela cells obtained using our DPC approach. (a) The image captured by the polarization camera.(b) Qualitative phase contrast image. (c) Quantitative phase image.

We conducted experimental demonstrations on unstained Hela cells using our DPC approach, and the results are shown in Fig.4. We can see that the qualitative phase contrast image (Fig. 4(b)) has higher contrast than the image captured by the polarization camera (Fig. 4(a)). The quantitative phase image (Fig. 4(c)) shows the internal details of the Hela cell Compared to Fig. 4(b).

In conclusion, we proposed a single-shot quantitative DPC method. By introducing polarizers in the illumination and imaging module of the system, we can decouple the two sets of asymmetric illumination acquisition images required to calculate the phase in a single acquisition. Most of the emerged single-shot DPC methods are wavelength multiplexing. Still, they need to introduce color-leakage correction methods to reduce the color crosstalk between the various wavelength channels that cause phase errors, as color LEDs tend to have broad-spectrum characteristics. Our polarization multiplexing strategy uses polarizing films of a higher extinction ratio directly over the LED, greatly reducing the channel crosstalk. The polarization camera can also provide accurate polarization data for each polarization channel. In addition, our approach uses a programmable LED array covered with polarizing films as illumination, and the system is simple, low-cost, and can be easily applied to existing commercial microscopes. The imaging performance of the method can be further improved later. For example, the computational correction of optical aberration can be considered in the decoupled two sets of asymmetric illumination acquisition images to obtain a phase reconstruction with digital aberration correction.

**Funding.** National Natural Science Foundation of China (62275020)

**Disclosures.** The authors declare no conflicts of interest.